\documentclass[12pt,preprint]{aastex}

\shorttitle{DNC/HNC RATIO in HMSFRs}
\shortauthors{Sakai et al.}

\begin{document}
%\baselineskip = 10mm


\title{DNC/HNC RATIO OF MASSIVE CLUMPS IN EARLY EVOLUTIONARY STAGES OF HIGH-MASS STAR FORMATION}


\author{Takeshi Sakai\altaffilmark{1,2}, Nami Sakai\altaffilmark{3}, Kenji Furuya\altaffilmark{4}, Yuri Aikawa\altaffilmark{4}, Tomoya Hirota\altaffilmark{2} and Satoshi Yamamoto\altaffilmark{3}}

\altaffiltext{1}{Institute of Astronomy, The Universityof Tokyo, Osawa, Mitaka, Tokyo 181-0015, Japan}
\altaffiltext{2}{National Astronomical Observatory of Japan, Osawa, Mitaka, Tokyo 181-8588, Japan.}
\altaffiltext{3}{Department of Physics, Graduate School of Science, The University of Tokyo, Tokyo 113-0033, Japan.}
\altaffiltext{4}{Department of Earth and Planetary Sciences, Kobe University, Kobe 657-8501, Japan.}

\begin{abstract}
We have observed the HN$^{13}$C $J$=1--0 and DNC $J$=1--0 lines toward 18 massive clumps, including infrared dark clouds (IRDCs) and high-mass protostellar objects (HMPOs), by using the Nobeyama Radio Observatory 45 m telescope.
We have found that the HN$^{13}$C emission is stronger than the DNC emission toward all the observed sources.
The averaged DNC/HNC ratio is indeed lower toward the observed high-mass sources (0.009$\pm$0.005) than toward the low-mass starless and star-forming cores (0.06).
The kinetic temperature derived from the NH$_3$ ($J$, $K$) = (1, 1) and (2, 2) line intensities is higher toward the observed high-mass sources than toward the low-mass cores.
However the DNC/HNC ratio of some IRDCs involving the Spitzer 24 $\mu$m sources is found to be lower than that of HMPOs, although the kinetic temperature of the IRDCs is lower than that of the HMPOs.
This implies that the DNC/HNC ratio does not depend only on the current kinetic temperature.
With the aid of chemical model simulations, we discuss how the DNC/HNC ratio decreases after the birth of protostars.
We suggest that the DNC/HNC ratio in star-forming cores depends on the physical conditions and history in their starless-core phase, such as its duration time and the gas kinetic temperature. 
\end{abstract}


\section{Introduction}

One of key issues toward a full understanding of the formation process of high-mass stars is to reveal their initial conditions.
For this purpose, infrared dark clouds (IRDCs) are thought to be good targets.
IRDCs are extinction features against the background mid-IR emission (e.g. P\'{e}rault et al. 1996; Egan et al. 1998).
More than ten thousands of IRDCs have been identified on the MSX data (Simon et al. 2006) and the Spitzer data (Peretto \& Fuller 2009).  
By mapping observations with bolometer arrays, many massive clumps ($>$100 $M_\odot$) have been found in IRDCs (e.g. Rathborne et al. 2006).
These massive clumps are known to be in various evolutionary stages of star formation. 
Chambers et al. (2009) classified the massive clumps associated with IRDCs into 4 groups on the basis of the 1.2 mm continuum data and the infrared data.  According to their results, star formation has already occurred in these massive clumps, except for some candidates of high-mass starless clumps.

It has been well established that chemical composition is a useful indicator of evolutionary stages for low-mass starless cores (e.g. Suzuki et al. 1992; Benson et al. 1998).  We considered that such a chemical approach would also be useful for high-mass sources, and carried out systematic surveys of several molecular lines toward massive clumps associated with IRDCs (Sakai et al. 2008; 2010).  
As a result, we found that the CCS/N$_2$H$^+$ ratio of massive clumps is lower than that of young low-mass starless cores.  This indicates that most of the massive clumps are chemically evolved than the young low-mass starless cores, since the CCS/N$_2$H$^+$ ratio is known to become lower for more evolved cores (Benson et al. 1998).
In addition, we found that the SiO and CH$_3$OH abundances relative to H$^{13}$CO$^+$ are enhanced in several massive clumps.   
Since SiO and CH$_3$OH are known as "shock tracers", these results suggest that the effect of shocks is more significant in the IRDCs than in the HMPOs.  Such shocks would be caused by an interaction between outflows and dense gas.

Recently, Vasyunina et al. (2011) surveyed the lines of 13 molecules toward massive clumps in IRDCs. They suggest that the chemical composition of the massive clumps in IRDCs is similar to that of low-mass prestellar cores rather than that of HMPOs.
In addition, spectral line observations toward massive clumps have extensively been carried out for fundamental molecules (Teyssier et al. 2002; Pillai et al. 2006; Purcell et al. 2006; Ragan et al. 2006; Beuther \& Sridharan 2007; Leurini et al. 2007a).
 
In spite of the above efforts, the initial conditions of high-mass star formation have not been established well.
In this paper, we examine it on the basis of the deuterium fractionation in massive clumps.
In cold molecular clouds, the deuterium atom is known to be fractionated into molecules except for HD (i.e. the main reservoir of the deuterium atom) through the following exothermic isotope-exchange reactions;
\begin{equation}
{\rm H}_3^+ + {\rm HD} \rightarrow {\rm H}_2{\rm D}^+ + {\rm H}_2 + 230 \; {\rm K},
\end{equation}
\begin{equation}
{\rm CH}_3^+ + {\rm HD} \rightarrow {\rm CH}_2{\rm D}^+ + {\rm H}_2 + 370 \;  {\rm K},
\end{equation}
\begin{equation}
{\rm C}_2{\rm H}_2^+ + {\rm HD} \rightarrow {\rm C}_2{\rm HD}^+ + {\rm H}_2 + 550 \; {\rm K}.
\end{equation}
The H$_3^+$, CH$_3^+$, and C$_2$H$_2^+$ ions are related to the formation processes of various molecules, 
so that the deuterium fractionation is transferred to their daughter molecules. 
Since the backward reaction rates of the above reactions are sensitive to the gas kinetic temperature,
the equilibrium deuterium fractionation ratios of molecules depend on the temperature.
Furthermore, the deuterium fractionation is enhanced in the evolved stages of starless cores, because CO, which is a main destroyer of H$_2$D$^+$, is depleted onto dust grains below 20 K. 
Such an enhancement was indeed confirmed toward some low-mass starless cores (Caselli et al. 1999).
Therefore, the deuterium fractionation ratio is regarded as a good indicator to trace the evolved starless core phase. 
In order to investigate the starless phase of high-mass star formation, Fontani et al. (2011) focused on the N$_2$D$^+$ and N$_2$H$^+$ ratio. They observed the N$_2$D$^+$ ($J$=2--1) and N$_2$H$^+$ ($J$=1--0, 3--2) lines toward 27 massive clumps including IRDCs, high-mass protostellar objects, and ultra-compact HII regions, and suggested that the N$_2$D$^+$/N$_2$H$^+$ ratio is higher in high-mass starless core candidates.
Pillai et al. (2007) also observed the NH$_2$D/NH$_3$ ratio toward 32 pre/protocluster sources.
They found moderately high deuterium fractionation ratios (0.001--0.7) in the massive clumps.

In general, the deuterium fractionation can be affected by star formation activities.
Since the gas kinetic temperature around protostars becomes higher than that in cold cores, the equilibrium deuterium fractionation ratios are lower in star forming cores.
However, the equilibrium condition would hardly be satisfied in actual star forming regions for the neutral species. 
Rodgers \& Millar (1996) pointed out that the initial deuterium fractionation ratio remains over 10$^4$ yr after the onset of star formation.
If so, the deuterium fractionation ratios of "star forming" cores contain novel information on physical conditions and history of their starless core phase.
By observing the deuterium fractionation ratios of the neutral species in high-mass star forming cores, we would be able to know their initial conditions before the birth of high-mass protostars.

Recently, candidates of high-mass starless cores have been found (e.g. Beuther et al. 2010), whose detailed characterization will provide us with a clue to understand the necessary conditions for high-mass star formation.  However a limited number of such sources has been known so far in the Galaxy (e.g. Chambers et al. 2009).   Furthermore, it is not guaranteed that massive stars will certainly form in such sources in the future. 
Our approach is to derive the initial condition of high-mass star formation from the DNC/HNC ratio of "star-forming" cores, which is complimentary to the effort investigating the candidates of high-mass starless cores.  
Based on this motivation, we have observed the DNC and HN$^{13}$C lines toward IRDCs and high-mass protostellar objects (HMPOs; Sridharan et al. 2002; Beuther et al. 2002a). 
We choose DNC and HN$^{13}$C, because Hirota et al. (2001) observed the DNC/HNC ratio toward low-mass cores with the Nobeyama Radio Observatory (NRO) 45 m telescope.
The HNC line is sometimes optically thick, and hence we observed the HN$^{13}$C line, which is generally optically thin.
Hence, we can critically compare the high-mass data with the low-mass data.
In this paper, we also compare the observation results with the model calculation results, and discuss how the deuterium fractionation ratios vary after the onset of star formation.

\section{Observations}

We selected 18 sources from the source lists of previous survey observations (Sakai et al. 2008; 2010).
In order to investigate how the deuterium fractionation ratios vary with evolution, 
we selected the targets with three different evolutionary stages on the basis of the mid-infrared properties:
IRDCs without Spitzer sources, IRDCs with Spitzer sources, and HMPOs.
An equivalent evolutionary classification of IRDCs and HMPO is also given in Chambers et al. (2009).
All the selected IRDCs are dark in the MSX 8  $\mu$m data.
The IRDCs without Spitzer sources are dark even in the Spitzer 24 $\mu$m data, 
so that high-mass star formation has not yet occurred there.  Hence, they are thought to be youngest among the three sources.
In the IRDCs with Spitzer sources, the Spitzer 8 $\mu$m or 24 $\mu$m sources are associated, so that star formation have already started.
Since the SiO and CH$_3$OH lines are detected toward the IRDCs with Spitzer sources (Sakai et al. 2010), 
these sources should have outflows.

HMPOs are thought to be most evolved among the three categories.  Since HMPOs are luminous in infrared and 
weak in the cm band, HMPOs are thought to be in pre-ultracompact/compact HII region phase (Sridharan et al. 2002; Beuther et al. 2002a).  We have selected the HMPOs toward which relatively strong CH$_3$OH line is detected by Sakai et al. (2010).
This indicates that the significant amount of dense gas still surrounds the protostars.
Since we found in the previous paper that dense gas has already disappeared in some HMPOs, 
our selected HMPOs can be regarded as relatively young ones.

The masses of all the targets are larger than 100 $M_\odot$, and the distances are all less than 4.6 kpc. 
Note that the distance to G034.43+00.24 is recently found to be 1.6 kpc with the VLBI observations (Kurayama et al. 2011).
This is much closer than the previous estimation based on the rotation curve and the LSR velocity (3.5 kpc). 
The target sources are listed in Table \ref{tab:t1}.

The HN$^{13}$C $J$=1--0 and DNC $J$=1--0 lines (Table \ref{tab:t2}) were observed with the NRO 45 m telescope in 2009 March. 
They were simultaneously observed by using the 2 side-band SIS receiver, T100 (Nakajima et al. 2008). 
The half-power beam width is about $21^{\prime\prime}$ and $18^{\prime\prime}$ at 76 and 87 GHz, respectively. The main beam efficiency ($\eta_{MB}$) is 0.53 and  0.43 at 76 and 87 GHz, respectively.
We derived the main beam temperature ($T_{MB}$) from the antenna temperature ($T_a^*$) by using the main beam efficiency as $T_{MB}$ = $T_a^*$/$\eta_{MB}$.
For all the observations, acousto-optical radiospectrometers (AOSs) were employed as a backend. We used AOS-Hs, each of which has the bandwidth and the frequency resolution of 40 MHz and 37 kHz, respectively.
The telescope pointing was checked by observing the nearby SiO maser source every 1--2 hours, and was maintained to be better than $5^{\prime\prime}$. 
The line intensities were calibrated by the chopper wheel method. 
All the observations were carried out with the position switching mode.
The emission free regions in the Galactic Ring Survey $^{13}$CO $J$=1--0 data (Jackson et al. 2006) were employed as the OFF positions, as reported by Sakai et al. (2008).

\section{Results \& Analysis}

\subsection{Spectra}

\begin{figure}
\figurenum{1}
 \epsscale{.80}
     \plotone{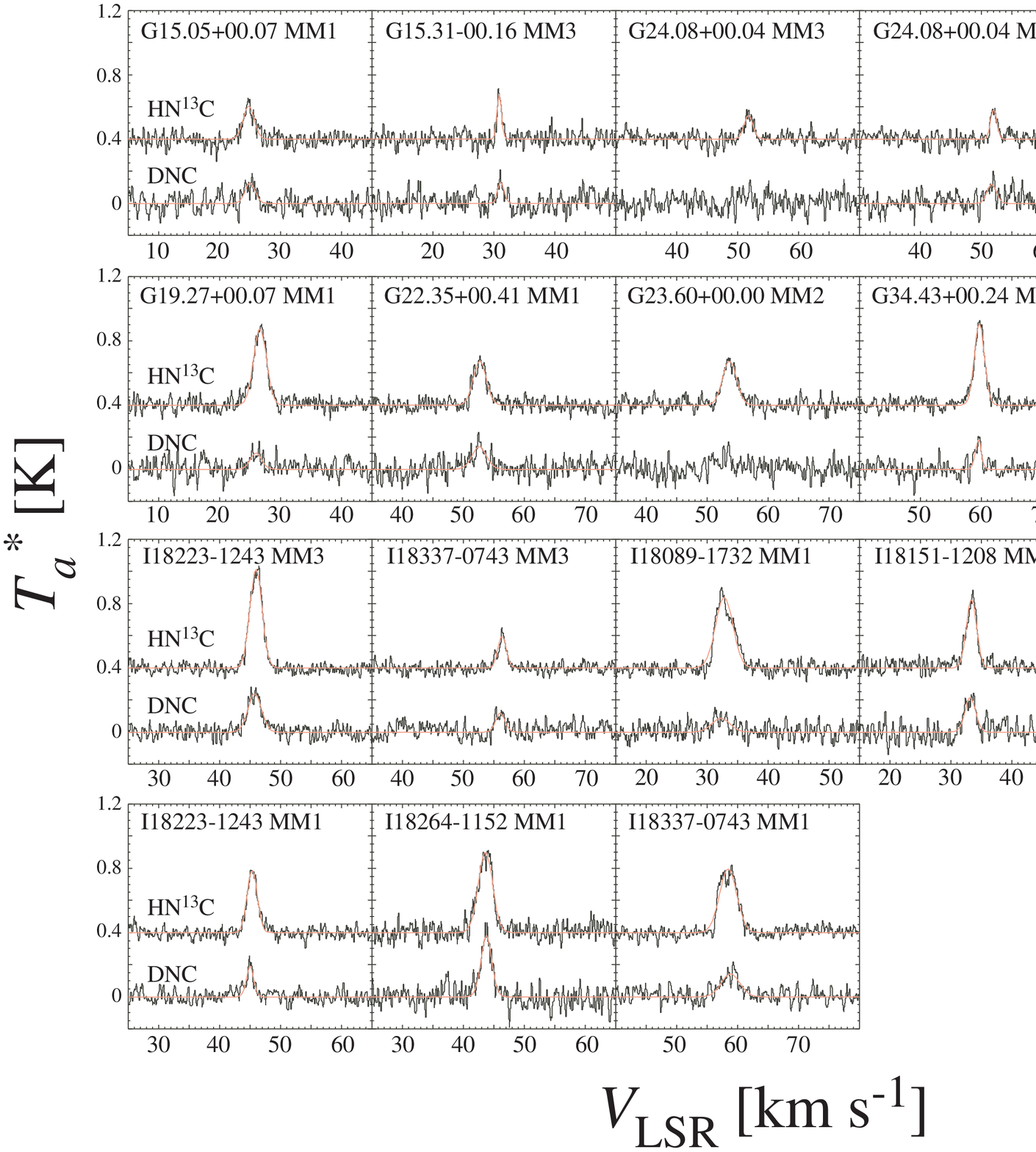}
	\caption{Spectral profiles of the HN$^{13}$C $J$=1--0 (top) and DNC $J$=1--0 (bottom) lines toward the observed sources. The fitting results are overlaid on the spectra as red lines.}
	\label{fig:f1}
\end{figure}

Figure \ref{fig:f1} shows the spectra of the observed sources.
We detected the HN$^{13}$C $J$=1--0 emission toward all the observed sources 
and the DNC $J$=1--0 emission toward 16 out of the 18 sources with the 3 $\sigma$ or higher confidence level at the velocity resolution of 0.08 km s$^{-1}$.
The DNC intensity appears to be weaker than the HN$^{13}$C intensity toward all the sources.
This is in contrast to the low-mass core cases, where the DNC intensity is stronger than the HN$^{13}$C intensity for most sources (Hirota et al. 2001).
Since the HN$^{13}$C and DNC $J$=1--0 lines are split into hyperfine components due to the nuclear quadrupole interactions of the N and D nuclei, we tried to fit the multiple Gaussian function to the observed spectral line profiles by considering hyperfine components, optical depth, and excitation temperature.
However, we could not get any reliable results about the optical depth and the excitation temperature for all the spectra, 
because the lines are almost optically thin.
Even if we fit the spectral line profiles to determine the optical depth and the velocity width by assuming the appropriate excitation temperature, the derived velocity widths are always very close to those found by the single Gaussian fit.
This is because their hyperfine splittings are much smaller than the typical velocity widths ($\sim$ 2 km s$^{-1}$) and the observed spectra show a single peaked profile, we fitted a single Gaussian function to all the spectra.
Although the spectral shape is sometimes different from the Gaussian shape, this is not due to the hyperfine structure, but is due to the velocity structure of the sources.
The derived line parameters of HN$^{13}$C and DNC $J$=1--0 are summarized in Tables \ref{tab:t3} and \ref{tab:t4}, respectively.

\begin{figure}
\figurenum{2}
 \epsscale{.90}
     \plotone{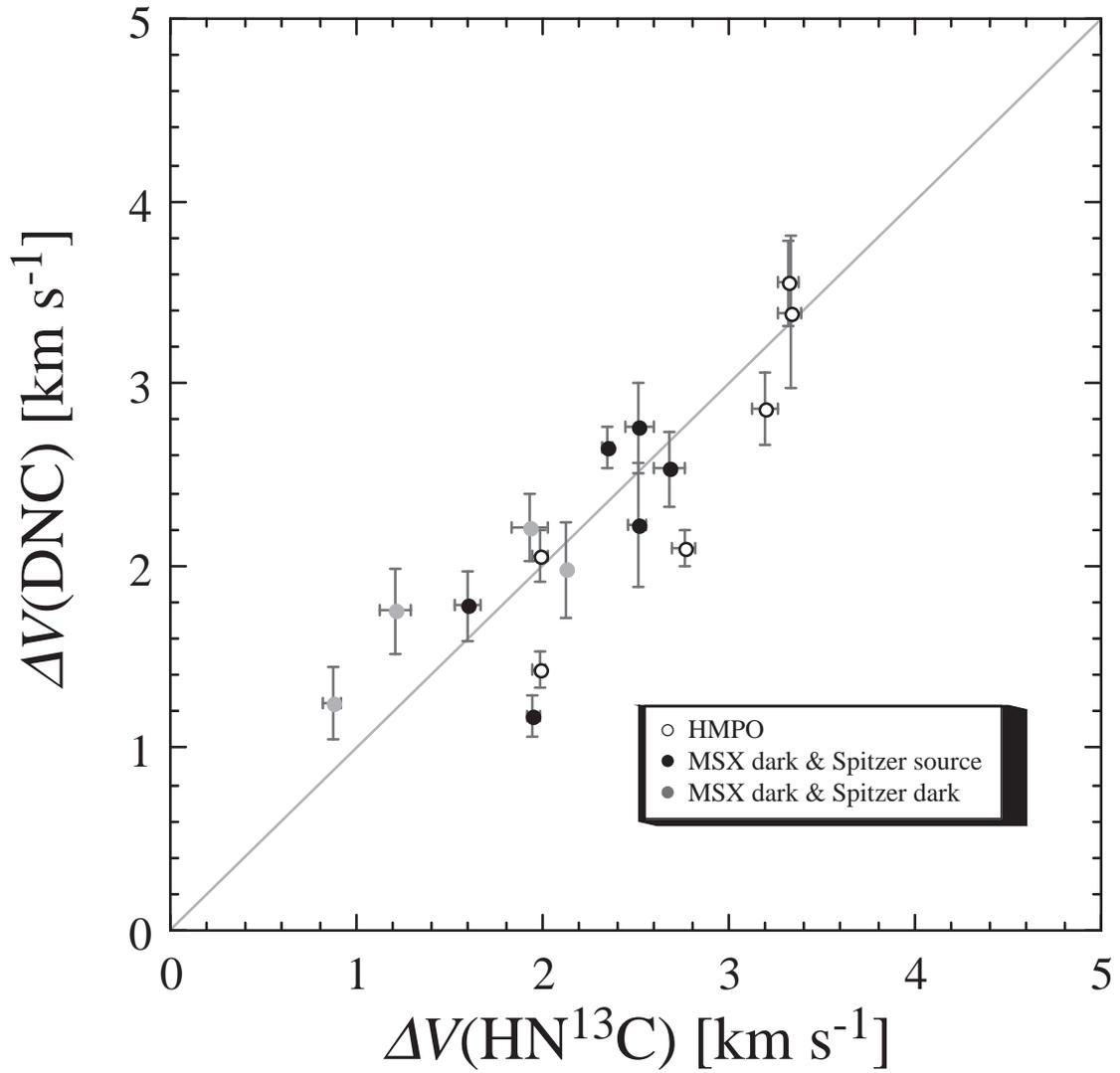}
	\caption{Plots of the velocity width of the DNC $J$=1--0 line against that of the HN$^{13}$C $J$=1--0 line. The grey line indicates the line of $\Delta V$(HN$^{13}$C) = $\Delta V$(DNC).}
	\label{fig:f2}
\end{figure}

Figure \ref{fig:f2} shows the plot of the HN$^{13}$C and DNC velocity widths.
A good correlation can be seen between them, where the correlation coefficient is 0.85. 
This indicates that the emitting regions of the both lines are similar to each other. 
Since the HN$^{13}$C emission comes from relatively quiescent regions, as suggested by Sakai et al. (2010), the DNC emission should also comes from such regions. Thus, the shock chemistry is not important for both HNC and DNC.

\subsection{Column Density Ratios}

We derive the column density ratio by assuming the local thermodynamic equilibrium condition.
The observed DNC and HN$^{13}$C lines are assumed to be optically thin, because their intensities are much lower than the gas kinetic temperature ($\sim$20 K) and their line shape is not flat-topped.
Since the HN$^{13}$C and DNC lines are likely to originate from the similar regions, as mentioned above, 
we assume the same excitation temperature and the same beam filling factor for the both lines.
In this case, the column density ratio can be expressed as
\begin{equation}
\frac{N({\rm DNC})}{N({\rm HN^{13}C})} =\left(\frac{\mu_{0, {\rm HN^{13}C}}}{\mu_{0, {\rm DNC}}}\right)^2 \left(\frac{\nu_{\rm HN^{13}C}}{\nu_{\rm DNC}}\right)^2 \exp\left(\frac{E_{u, {\rm DNC}} - E_{u, {\rm HN^{13}C}}}{kT_{\rm ex}}\right) \frac{1-\frac{J_{\rm HN^{13}C}(T_{\rm BB})}{J_{\rm HN^{13}C}(T_{\rm ex})}}{1-\frac{J_{\rm DNC}(T_{\rm BB})}{J_{\rm DNC}(T_{\rm ex})}}\frac{I_{\rm DNC}}{I_{\rm HN^{13}C}},
\end{equation}
where $N$ denotes the column density, $\mu_0$ the dipole moment, $\nu$ the transition frequency, $E_u$ the upper state energy, $T_{\rm ex}$ the excitation temperature, $T_{\rm BB}$ the cosmic background radiation temperature (2.7 K), and $I$ the integrated intensity in $T_{\rm MB}$. 
Furthermore, $J$($T$) stands for the radiation temperature given as
\begin{equation}
J(T) = \frac{\frac{h\nu}{k}}{\exp\left(\frac{h\nu}{kT}\right)-1}.
\end{equation}
When we use the values in Table \ref{tab:t2} for $\mu_0$, $\nu$ and $E_u$, the above equation is reduced as
\begin{equation}
\frac{N({\rm DNC})}{N({\rm HN^{13}C})} \simeq 1.30 \exp\left(-\frac{0.52}{T_{\rm ex}}\right) \frac{I_{\rm DNC}}{I_{\rm HN^{13}C}}.
\end{equation}

As for the excitation temperature, we adopt the rotation temperature of NH$_3$.
The NH$_3$ data were obtained with the NRO 45 m telescope by Sakai et al. (2008), except for two sources (I18089-1732 MM1 and I18264-1152 MM1).
For these two sources, we use the NH$_3$ data observed with the 100 m telescope by Sridharan et al. (2002).
The derived column density ratios are summarized in Table \ref{tab:t5}.
The errors denoted in Table \ref{tab:t5} include the uncertainty of the excitation temperature as well as the rms noise.
As evident from Equation (5), the ratio is not sensitive to the assumption of $T_{\rm ex}$, because ($E_{u.{\rm DNC}}$-$E_{u.{\rm HN^{13}C}}$)/$kT_{\rm ex}$ is close to 0.
Even when we change the excitation temperature from 10 to 100 K, the column density ratio varies only by 5 \%.
In Table \ref{tab:t5}, we also list $N$(HN$^{13}$C) as well as the rotation temperature of NH$_3$ adopted as the excitation temperature for reference.

We note that the column density ratio derived here is the average value within the observed beam.
This is the typical problem which we face in the single-dish observations of the sources at few kpc distance.
However, this does not cause a serious problem in our study.
Since the critical density is as high as 10$^6$ cm$^{-3}$ for both of the DNC and HN$^{13}$C $J$=1--0 lines,
the derived DNC/HNC ratio would reflect the properties of the high-density regions.
Such high-density regions are generally linked with star formation in the sources.
Therefore, we would be able to discuss the effect of star formation from the derived DNC/HNC ratio,
even if we do not spatially resolve possible internal structures within clumps.
The effect of limited angular resolution will also be discussed later (Section 4.1).

 To derive the column density of HNC,  we assume the $^{12}$C/$^{13}$C ratio given by the following equation: $^{12}$C/$^{13}$C = 7.5$D_{\rm GC}$ + 7.6 (Wilson \& Rood, 1994).   The $^{12}$C/$^{13}$C ratio used in the present analysis ranges from 40 to 62, whereas Hirota et al. (2001) used the $^{12}$C/$^{13}$C of 60. Then, the averaged value of the $N$(DNC)/$N$(HNC) ratio is evaluated to be 0.009 ($\pm$0.005).
This ratio is lower than the corresponding ratio of the low-mass star forming regions (Hirota et al. 2001).
Fontani et al. (2006) also suggest that the N$_2$D$^+$/N$_2$H$^+$ ratio for the high-mass protostellar candidates is lower than that for low-mass cores (see Figure 2 in Fontani et al. 2006). Our result for the DNC/HNC ratio is consistent with theirs.
As for high-mass star forming regions, the observed ratio is comparable to the $N$(N$_2$D$^+$)/$N$(N$_2$H$^+$) ratio ($\sim$0.015) reported by Fontani et al. (2006), while it is lower than the $N$(N$_2$D$^+$)/$N$(N$_2$H$^+$) ratio by Fontani et al. (2011; 0.02--0.7) and $N$(NH$_2$D)/$N$(NH$_3$) ratio by Pillai et al. (2007; 0.1--0.7).

\section{Discussion}

\subsection{Temperature Dependence of the DNC/HNC Ratio}

\subsubsection{Observational Trend}

As mentioned in Introduction, the equilibrium deuterium fractionation ratios would be affected by the gas kinetic temperature.
To examine this, we estimate the kinetic temperature from the NH$_3$ rotation temperature by using the equation 
(2) in Walmsley \& Ungerechts (1983). For discussion, we only use the NH$_3$ data obtained by Sakai et al. (2008) in order to minimize the systematic errors caused by the difference in angular resolution.
We exclude I18089-1732 MM1 and I18264-1152 MM1 in the following discussion, because we do not have the NH3 data observed with the NRO 45 m telescope. 
Although the NH$_3$ lines were observed toward these objects by Sridharan et al. (2002) using the Effelsberg 100 m telescope, the beam size (40$^{\prime\prime}$) is smaller than that of our NRO 45 m 
observations by a factor of 2.  
The kinetic temperature of I18089-1732 MM1 is much higher than the other sources. 
This may be partly related to the difference in the angular resolution.

In Figure \ref{fig:f3}, we plot the $N$(DNC)/$N$(HNC) ratio against the gas kinetic temperature for the observed sources.
For comparison, we also plot the data of low-mass cores, which are reported by Hirota et al. (2001). 
We only use the data of the low-mass cores toward which the gas kinetic temperature is reported (Jijina et al. 1999 and references therein).
The temperature is systematically higher in the high-mass sources than in the low-mass sources.
In Figure \ref{fig:f3}, we can clearly see the difference in the DNC/HNC ratio between low- and high-mass star forming regions.  
The average value of the DNC/HNC ratio is 0.009 ($\pm$0.005) for the observed high-mass sources, while it is 0.06 ($\pm$0.03) for the low-mass sources.

Since the high-mass sources are generally more distant than the low-mass sources, the lower DNC/HNC ratio might be due to the lower spatial resolution for the high-mass sources.  However, Hirota et al. (2003) revealed the distribution of the DNC/HNC ratio toward the low-mass cores, and found that the DNC/HNC ratio is quite uniform within the cores with a scale of about 3$^{\prime}$ from the edge to the center.
Even if the low-mass cores would be located at a distance of a few kpc, which is more distant than the low-mass cores typically by a factor of 10, the high DNC/HNC ratio would be observed with our 20$^{\prime\prime}$. 
We therefore think that the lower DNC/HNC ratios of the high-mass sources are not due to the distance effect.
Indeed, there is no correlation between the distance and the DNC/HNC ratio.

In addition, we can see a trend that the $N$(DNC)/$N$(HNC) ratios of the Spitzer-dark sources are slightly higher than those of the other sources.
The average value for the Spitzer-dark sources is also higher than that for the other sources (see Table \ref{tab:t5}).
This trend is similar to that reported by Fontani et al. (2011), where the $N$(N$_2$D$^+$)/$N$(N$_2$H$^+$) ratios of high-mass starless cores tend to be higher than that of the HMPOs and UC HII regions.
However, our observed Spitzer dark sources have lower $N$(DNC)/$N$(HNC) ratios than the low-mass cores. 
This is different from that found by Fontani et al. (2011), where the $N$(N$_2$D$^+$)/$N$(N$_2$H$^+$) ratio of several HMSCs is comparable to that of low-mass starless cores.
This difference may partly be due to a diversity of the deuterium fractionation in the high-mass starless core phase.
Indeed, the $N$(N$_2$D$^+$)/$N$(N$_2$H$^+$) ratios of some high-mass starless cores are as low as those of the HMPOs and the UC HII regions even in the sample of Fontani et al. (2011).
In addition, Tatematsu et al. (2010) found a starless core that has the low DNC/HNC ratio ($\sim$0.008) toward the Orion A giant molecular cloud.
On the other hand, the difference may be attributed to chemical reasons.  If N$_2$D$^+$ and DNC is formed mainly from H$_2$D$^+$ and CH$_2$D$^+$, respectively (Roueff et al. 2007), for instance, the deuterium fractionation ratio is thought to be different between N$_2$D$^+$ and DNC.  We need to observe every sources with the same molecular lines in order to confirm the diversity of the deuterium fractionation ratio in the starless phase.

In the present stage, it is still controversial whether the Spitzer-dark sources are really in the starless phase of high-mass stars.
The HN$^{13}$C column density of our observed Spitzer-dark sources is smaller than that of the other sources (Table \ref{tab:t5}), and is comparable to that of the low-mass cores (Hirota et al. 2001).  In addition, the 1.2 mm peak flux of the Spitzer-dark sources (37 to 115 mJy) is lower than that of the Spitzer-bright sources (110 to 424 mJy) (Beuther et al. 2002; Rathborne et al. 2006).
It may be possible that the lower HNC abundance is due to the depletion onto dust grains, and that the lower mm-continuum emission is due to the low temperature.
Nevertheless, we cannot rule out a possibility that the observed Spitzer dark sources are smaller than the other sources and they are just low-mass star forming regions.
On the other hand, the HNC column density is comparable between the Spitzer-bright IRDCs and HMPOs, and hence,
we may say that their masses are similar.
Thus, we think that comparison of the DNC/HNC ratio between the Spitzer-bright IRDCs and the HMPOs is much more secure to investigate the evolution of the DNC/HNC ratio.

\subsubsection{Comparison with the Chemical Model}

\begin{figure}
\figurenum{3}
 \epsscale{.80}
     \plotone{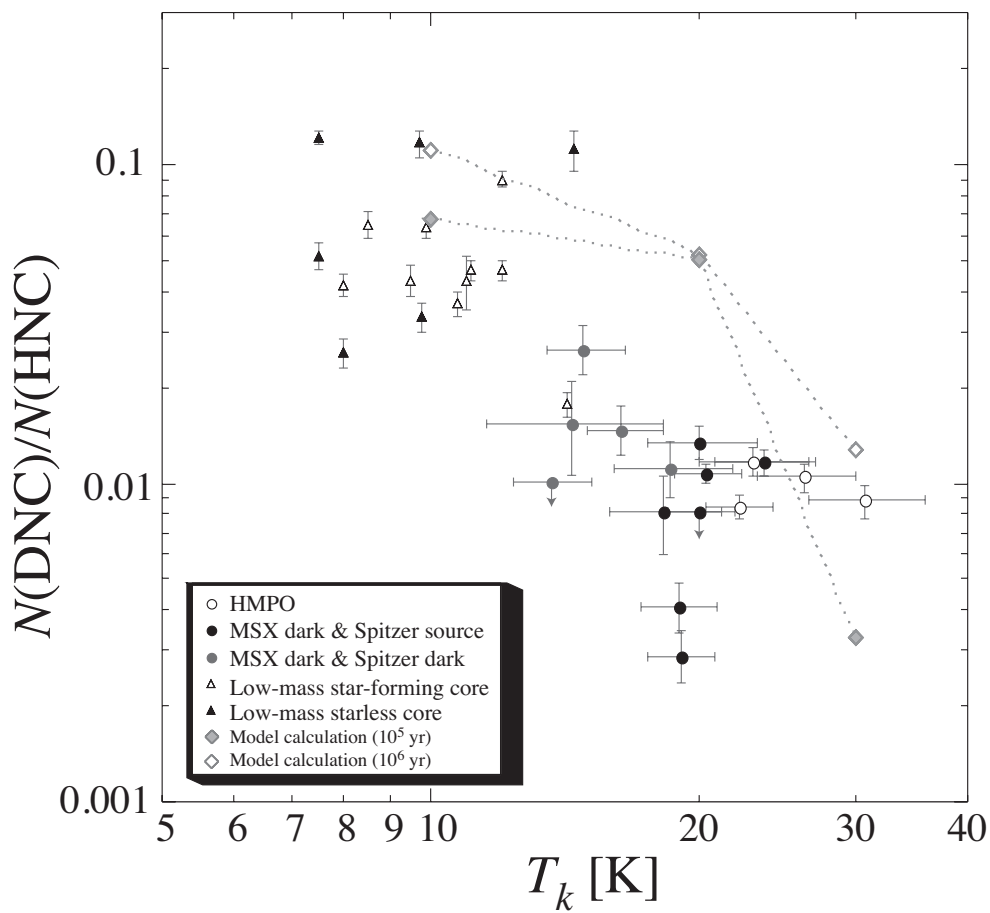}
	\caption{Plot of the $N$(DNC)/$N$(HNC) ratio against the gas kinetic temperature derived from the NH$_3$ ($J$, $K$) = (1, 1) and (2, 2) lines. The low-mass data are taken from Hirota et al. (2001). The $N$(DNC)/$N$(HNC) ratio derived from the chemical model calculation is also shown for the ages of 1$\times$10$^5$ yr or 3$\times$10$^5$ yr.}
	\label{fig:f3}
\end{figure}

To investigate whether the lower DNC/HNC ratio of the high-mass sources is explained only by the difference of the gas kinetic temperature, we performed chemical model calculations.
The chemical network used is the same as that reported by Aikawa et al. (2005).
This model includes the gas-phase reactions, grain-surface reactions, depletion onto dust grains, and desorption from dust grains.
We assume a static geometry with a constant H$_2$ density of 10$^5$ cm$^{-2}$ for simplicity, and calculated the temporal variation of the molecular abundances for 10 K, 20 K, and 30 K.  
We assume the so-called low-metal values  (see Table 1 of Aikawa et al. 2001) as the elemental abundances.
All the heavy elements are assumed to be initially in the atomic form and the hydrogen in the molecular form.
The deuterium is assumed to be in the form of HD, with the fractional abundance of 3.0$\times$10$^{-5}$ relative to H$_2$.
We adopt the binding energy of CO on the dust to be 1180 K.  Therefore, CO still remains on the dust at the temperature of 20 K, and the sublimation temperature of CO is about 21 K in this model. 
The results at 10$^5$ yr and 10$^6$ yr after starting chemical reactions from the initial condition are plotted in Figure \ref{fig:f3}.

According to the model calculation, the DNC/HNC ratio does not change very much between 10 K and 20 K, while it considerably decreases at 30 K. 
Roueff et al. (2007) also reported a similar result for DNC/HNC. 
In Figure \ref{fig:f3}, the observed DNC/HNC ratios for the low-mass sources are consistent with the model calculation. 
On the other hand, the ratios observed for the massive clumps are lower than the model prediction.
The discrepancy is particularly significant for the sources with the gas kinetic temperature around 20 K.

Roueff et al. (2007) predicted that the D/H ratio for some molecules such as C$_2$H$_2^+$ could become, in general, higher in denser regions. 
However, they suggest from their model calculation that the equilibrium DNC/HNC ratio does not depend on the density in the low-metal abundance case.
Therefore, the differences of the H$_2$ density would not explain the trend seen in Figure \ref{fig:f3}.
Indeed, it is unlikely that the high-mass star forming regions are less dense than the low-mass star forming regions.

It is known that existence of the ortho H$_2$ molecule enhances the backward reaction of (1), because the ortho H$_2$ molecule in its lowest rotational state ($J$=1) has the rotational energy of about 170 K (Gerlich et al. 2002; Flower et al. 2006).  Hence, it decreases the deuterium fractionation ratio of H$_2$D$^+$ and those of various molecules.  We ignored this effect in the chemical model calculation, because the ortho-to-para ratio of H$_2$ is reported to be close to 0 for low-mass protostar sources (Flower et al. 2006; Caselli et al. 2008; Pagani et al. 2009; Troscompt et al. 2009).  If the ortho-to-para ratio of H$_2$ is significantly higher in the massive clumps, the lower deuterium fractionation at 20 K could be explained.  However, the fractionation ratio at higher temperatures would also have to decrease in this case, which contradicts with the ratios in HMPOs (Figure 3).  Although the ortho-to-para ratio of H$_2$ may affect the deuterium fractionation ratio to some extent, we do not consider it in the following discussion for simplicity.

Another possibility for the discrepancy between the model and observational results is that the DNC and HN$^{13}$C emissions would come from warmer regions than the NH$_3$ (1, 1) and (2, 2) emissions.
Since the critical density of the DNC and HN$^{13}$C lines ($\sim$10$^6$ cm$^{-3}$) is higher than that of the NH$_3$ lines ($\sim$10$^4$ cm$^{-3}$), the DNC and HN$^{13}$C emissions trace denser regions.
In the objects associated with the Spitzer sources, the inner dense part is likely to be warmer due to various activities of newly born stars, and hence, the DNC and HNC emissions tend to trace such regions. 
The smaller beams size of the DNC and HNC observations in comparison with the NH$_3$ observations may also contribute to this effect.
If this is the case, the temperature estimated from the NH$_3$ observation may be underestimated, and the data points around 20 K may have to be shifted to the higher temperature, giving a better agreement with the chemical model calculation.

However, some massive clumps have relatively lower DNC/HNC ratio than the HMPOs, although the gas kinetic temperature of the massive clumps is lower than that of the HMPOs (Figure \ref{fig:f3}).
For example, the DNC/HNC ratio of G034.43+00.24 MM3 (0.0029) is lower than that of I18151-1208 MM1 (0.0088), while the kinetic temperature of G034.43+00.24 MM3 (19 K) is lower than that of I18151-1208 MM1 (31 K). 
This trend is not explained only with the difference in the gas kinetic temperature.

In Figure \ref{fig:f3}, the model results at 10$^5$ and 10$^6$ yr are shown. 
At 10$^5$ yr, the deuterium fractionation ratio does not reach its steady state value, as seen in the difference of the ratio between 10$^5$ and 10$^6$ yr.
This means that the deuterium fractionation ratio depends not only on the current temperature but also on the past physical history of the clump.
Hence, the time dependence effect has to be considered for the DNC/HNC ratio, as discussed in the next section.

\subsection{Time dependence of the DNC/HNC ratio}

\subsubsection{Basic Mechanisms for Variation of the Molecular D/H Ratios after the Onset of Star Formation}

When we consider the evolution of molecular clouds from diffuse clouds into dense cores, the deuterium fractionation ratio of molecules gradually increases with time from the cosmic D/H ratio ($\sim$10$^{-5}$); the deuterium fractionation is expected to be low in the early stage of molecular cloud evolution.
In fact, Hirota et al. (2001) found the low deuterium fractionation in a few young low-mass starless cores.
In addition, Crapsi et al. (2005) found that the deuterium fractionation ratio is higher in the cores closest to the onset of gravitational collapse among the sample of low-mass starless cores, and they suggest that the deuterium fractionation increases with the core evolution.

The birth of protostars raises the temperature around them.
The deuterium fractionation ratio then starts to decrease toward the equilibrium value at the new temperature.
The timescale of this change is different from species to species, depending on the destruction timescale of each species. 
First, we consider the timescale for DNC/HNC.
The main destruction pathways of DNC are as follows;
\begin{equation}
{\rm DNC} + {\rm HCO^+} \rightarrow {\rm DNCH^+} + {\rm CO},
\end{equation}
\begin{equation}
{\rm DNC} + {\rm H_3^+} \rightarrow {\rm DNCH^+} + {\rm H_2}.
\end{equation}
Although the dissociative recombination of DNCH$^+$ can reproduce DNC, it competes with HCN + D
and CN + D + H.  
Thus, the abundance of DNC could decrease with the above reactions.
The timescale of the above reactions (7) and (8) is comparable to each other, and depends on the abundances of HCO$^+$ and H$_3^+$.
For instance, the timescale of the reaction with HCO$^+$ is represented as
\begin{equation}
\tau \sim \frac{1}{k[{\rm HCO^+}]} \sim 4 \times 10^{12} \; {\rm s} \sim 10^5 \; {\rm yr},
\end{equation}
in the case of the density of 10$^5$ cm$^{-2}$. Since the HCO$^+$ abundance depends on the H$_2$ density  (Umebayashi \& Nakano, 1980), the above timescale would have the weak power-low dependence on the density as $\tau \propto n^{0.3\sim 0.5}$.
Therefore, the DNC/HNC ratio can be preserved after the temperature rise.

On the other hand, we note that the destruction timescale of molecular ions is much shorter than that of neutral molecules, because they can react with CO and/or electron.
For example, the main destruction pathway of N$_2$D$^+$ is the following reaction;
\begin{equation}
{\rm N}_2{\rm D}^+ + {\rm CO} \rightarrow {\rm N_2} + {\rm DCO}^+.
\end{equation}
The destruction timescale of N$_2$D$^+$ is therefore given as
\begin{equation}
\tau \sim \frac{1}{k[{\rm CO}]} \sim 10^{8} \; {\rm s} \sim 4 \; {\rm yr} .
\end{equation}
The deuterium fractionation ratio of molecular ions immediately decreases after the onset of star formation, and reaches at the equilibrium value at the new temperature.

\subsubsection{Chemical Model Calculation}

\begin{figure}
\figurenum{4}
 \epsscale{1.0}
     \plotone{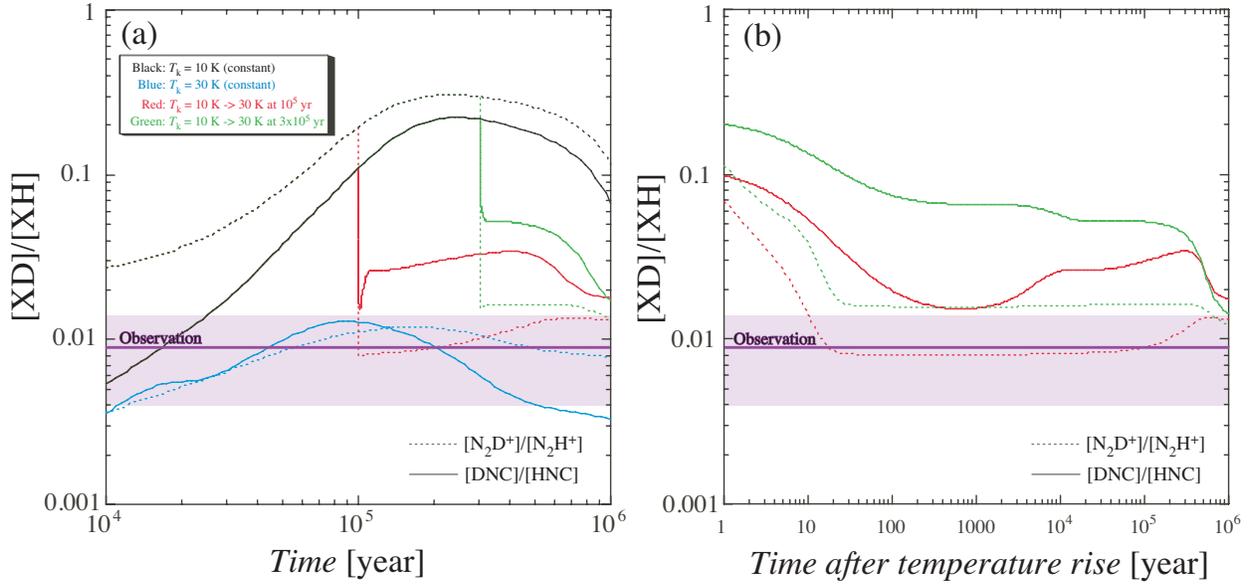}
	\caption{(a) Chemical model calculation of the deuterium fractionation ratio of DNC/HNC and N$_2$D$^+$/N$_2$H$^+$. 
The average value of the observation data for the high-mass sources (0.009) is plotted as a purple line, and one standard deviation of the average value (0.005) is indicated as a purple area. (b) Same as (a), but the time is measured from the time when the temperature is changed from 10 K to 30 K. See the text for details.}
	\label{fig:f4}
\end{figure}

To investigate how the deuterium fractionation ratio changes after the temperature rise, we performed the following chemical model calculations.
We simply assume a sudden temperature rise from 10 to 30 K at a given age, 1$\times$10$^5$ yr or 3$\times$10$^5$ yr, and calculate the temporal variation of chemical compositions after that.
We plot the calculated DNC/HNC and N$_2$D$^+$/N$_2$H$^+$ ratios in Figure \ref{fig:f4}a.
We also plot the DNC/HNC and N$_2$D$^+$/N$_2$H$^+$ ratios for the cases without the temperature rise.  

As seen in Figure \ref{fig:f4}a, the DNC/HNC and  N$_2$D$^+$/N$_2$H$^+$ ratios slowly increases with time until the temperature rise, and they decrease after that.
The decrease of the deuterium fractionation ratios originates from the above gas phase processes.
Although molecules are evaporated from dust grains after the temperature rise,
the DNC/HNC ratio in the gas phase is not affected by this effect.
HNC is not efficiently formed by grain-surface reactions, so that the DNC/HNC ratio in ice mantle is not significantly different from the ratio in the gas phase.

For clarity, we plot the DNC/HNC ratio after the temperature rise in Figure \ref{fig:f4}b.
In Figure \ref{fig:f4}b, we can see that the DNC/HNC ratio gradually decreases with a timescale of several times 10$^4$ yr,
while the N$_2$H$^+$ ratio rapidly decreases with a timescale of several 10 yr. 
This is consistent with the rough estimate of the timescales mentioned before.
More importantly, the DNC/HNC ratio after the temperature rise depends on the time when the temperature is raised.
In other words, the DNC/HNC ratio in the star forming clumps depends on the initial DNC/HNC ratio before the onset of star formation.
Since the dynamical age of outflows of the HMPOs are estimated to be several times 10$^4$--10$^5$ yr (Beuther et al. 2002b), the initial DNC/HNC ratio is conserved in the massive clumps and HMPOs to some extent. 
As mentioned in Section 4.1, some massive clumps with the Spitzer sources have lower DNC/HNC ratio than the HMPOs.
This difference would reflect the difference of the deuterium fractionation ratio at the onset of star formation.

For comparison, we plot the average value of all of our observed sources in Figure \ref{fig:f4}.
The observed value is comparable to the constant temperature case of 30 K.
On the other hand, the cases that the temperature rise from 10 K to 30 K at 10$^5$ yr and 3$\times$10$^5$ yr cannot reproduce the observed value, because the DNC/HNC ratio just before the temperature rise is preserved after the temperature rise.
This indicates that the DNC/HNC ratio just before the onset of star formation should have been low for these sources.
Thus, we speculate that the DNC/HNC ratio just before the onset of star formation would be lower for the observed high-mass sources than for the low-mass cores.
The lower DNC/HNC ratio implies that the cold starless phase of the observed sources is shorter than that of the low-mass cores, or that the initial temperature is warmer ($>$ 20 K) than that of the low-mass cores ($\sim$10 K).

The timescale of high-mass starless phase is still controversial.  From the statistical study, Chambers et al. (2009) estimated the timescale of starless phase to be 3.7$\times$10$^5$ yr, while Parsons et al. (2009) estimated it to be a few 10$^3$-10$^4$ yr.
Although our observed results seem to support the short timescale of the starless phase, we need more careful considerations.
In fact, Sakai et al. (2008) measured the CCS/N$_2$H$^+$ ratio toward the IRDCs and HMPOs, and obtained only the upper limits of the CCS/N$_2$H$^+$ ratio.
From these results, they suggest that massive clumps in IRDCs and HMPOs are chemically more evolved than young low-mass dense cores.
However, this argument is based on the single-dish observations, which may be affected by the limited angular resolution and sensitivity.
If the DNC/HNC ratio reflects the timescale of starless core phase, we should see a correlation between the DNC/HNC ratio and the CCS/N$_2$H$^+$ ratio with high-sensitivity and high-angular resolution observations.  
In this relation, Devine et al. (2011) have recently found the chemically young envelope toward the IRDC G19.30+00.07 with the VLA observations.   Existence of such a chemically young envelope may be related to the chemical youth of the clump.

The initial temperature of high-mass star formation has not been established well either.
The theoretical models of high-mass star formation generally assume the initial temperature from 10 to 30 K (e.g. Krumholz et al. 2009; Peters et al. 2011).
However, this assumption is not based on the observational results.
If the deuterium fractionation ratio would reflect the initial temperature, it will provide us with novel information for understanding the high-mass star formation, because the initial temperature affects the accretion rate and the gravitational instability causing clump fragmentation.

Discrimination of these two possibilities is not an easy task. 
A systematic observation of N$_2$D$^+$/N$_2$H$^+$ toward the same sources would be an important clue to this, as shown in Figure \ref{fig:f4}a.
The N$_2$D$^+$/N$_2$H$^+$ ratio is expected to be higher than the DNC/HNC ratio, if the clump temperature is 30 K from the starless-core phase (See the curves for the constant temperature of 30 K).

For a future work, we are going to observe the DNC/HNC and N$_2$D$^+$/N$_2$H$^+$ ratios toward an IRDC clump 
with the high-angular resolution by using ALMA.  From this observation, we will obtain important information
about the behavior of the deuterium fractionation in the high-mass star forming regions in more detail.

\section{Summary}

We conducted a survey of the HN$^{13}$C $J$=1--0 and DNC $J$=1--0 lines toward the 18 massive clumps, and compared the results with the chemical model calculation.
The main results are summarized as follows.

\begin{itemize} 

\item We detected the HN$^{13}$C $J$=1--0 emission toward all the observed sources 
and the DNC $J$=1--0 emission toward 16 out of the 18 sources above the 3 $\sigma$ confidence level at a velocity resolution of 0.08 km s$^{-1}$.
The intensities of the DNC line appear to be weaker than those of the HN$^{13}$C line toward all the sources in contrast to the case for the low-mass star forming regions.

\item The DNC/HNC ratio averaged for the observed sources except for the two sources is found to be 0.009 ($\pm$0.005), which is lower than that of the low-mass cores.
The kinetic temperature derived from the NH$_3$ ($J$, $K$) = (1, 1) and (2, 2) line intensities is higher toward the observed high-mass sources than toward the low-mass cores.
However, the ratios observed for the massive clumps are lower than the model calculation, particularly for the sources with the gas kinetic temperature around 20 K. Furthermore, we found that the DNC/HNC ratios of some massive clumps with the Spitzer sources are lower than that of HMPOs.
These results suggest that the DNC/HNC ratio cannot simply be explained by the current temperature of the clumps.

\item From the model calculations, it is found that the DNC/HNC ratio decreases with a timescale of several times 10$^4$ yr after the birth of protostars, whereas the N$_2$D$^+$/N$_2$H$^+$ ratio decreases with a timescale of less than 100 yr. Therefore, the DNC/HNC ratio of star forming clumps can reflect the initial DNC/HNC ratio just before the birth of protostars.  

\end{itemize}

\acknowledgments
We are grateful to the NRO staff for excellent support in the 45 m observations.
The 45 m radio telescope is operated by Nobeyama Radio Observatory, a branch of National Astronomical Observatory of Japan.
This study is supported by Grant-in-Aid from Ministry of Education, Culture, Sports, Science, and Technologies (21740144, 21224002, 23740146 and 23540266).




\clearpage


\clearpage

\clearpage

\clearpage

\clearpage

\begin{deluxetable}{lrrrrrr}
\tablecolumns{7} 
\tablewidth{0pc} 
\tabletypesize{\small} 
\tablecaption{Target source list.} 
\tablehead{ 
\colhead{Source} & \colhead{R. A.}   & \colhead{Dec.}    & \colhead{$V_{\rm LSR}$}   & \colhead{$D$\tablenotemark{a}}  & \colhead{$D_{\rm GC}$\tablenotemark{b}} &\colhead{Reference}\\
 & \footnotesize(J2000.0) & \footnotesize(J2000.0)  & \footnotesize[km s$^{-1}$] &  \footnotesize[kpc] &  \footnotesize[kpc] & \\
 }
\startdata 
\multicolumn{7}{c}{MSX \& Spitzer Dark Sources}\\ \hline
G015.05+00.07 MM1 & 18 17 50.4 & -15 53 38 &  24.7 & 2.5 &6.1& 1\\
G015.31-00.16 MM3 & 18 18 45.3 & -15 41 58 & 30.8 & 3.0 &5.6& 1\\
G024.08+00.04 MM3 & 18 35 02.2 & -07 45 25 & 51.9 & 3.6 &5.5& 1\\
G024.08+00.04 MM4 & 18 35 02.6 & -07 45 56 &  52.0 & 3.6 &5.4& 1\\
G034.43+00.24 MM9 & 18 53 18.4 & 01 28 14 & 59.0 & 1.6\tablenotemark{c} &7.3\tablenotemark{c}& 1\\ \hline
\multicolumn{7}{c}{MSX Dark Sources with Spitzer Sources}\\ \hline
G019.27+00.07 MM1 & 18 25 58.5 & -12 03 59 & 26.8 & 2.3 &6.4& 1\\
G022.35+00.41 MM1 & 18 30 24.4 & -09 10 34 & 52.7 & 3.7 &5.2& 1\\
G023.60+00.00 MM2 & 18 34 21.1 & -08 18 07 & 53.3 & 3.7 &5.3& 1\\
G034.43+00.24 MM3 & 18 53 20.4 & 01 28 23 & 59.7& 1.6\tablenotemark{c} &7.3\tablenotemark{c}& 1\\
I18151-1208 MM2 & 18 17 50.4 & -12 07 55 & 29.7 & 2.6 &6.1& 2,3,4\\
I18223-1243 MM3 & 18 25 08.3 & -12 45 28 & 45.7& 3.7 &5.2& 2,3,4\\
I18337-0743 MM3 & 18 36 18.2 & -07 41 01 & 56.4& 3.8 &5.3& 2,3,4\\ \hline
\multicolumn{7}{c}{High-Mass Protostellar Objects}\\ \hline
I18089-1732 MM1 & 18 11 51.5 & -17 31 29 & 33.1 & 3.6 &5.1& 2,3,4\\
I18151-1208 MM1 & 18 17 58.0 & -12 07 27 & 33.2 & 2.8 &5.9& 2,3,4\\
I18182-1433 MM1 & 18 21 09.2 & -14 31 57 & 60.0 & 4.6 &4.3& 2,3,4\\
I18223-1243 MM1 & 18 25 10.5 & -12 42 26 &45.3 & 3.6 &5.2& 2,3,4\\
I18264-1152 MM1 & 18 29 14.6 & -11 50 22 &43.7 & 3.4 &5.4& 2,3,4\\
I18337-0743 MM1 & 18 36 41.0 & -07 39 20 &58.4 & 3.9 &5.2& 2,3,4\\
\enddata 
\tablerefs{(1) Rathborne et al. (2006); (2) Sridharan et al. (2002); (3) Beuther et al. (2002a); (4) Sridharan et al. (2005).}
\tablenotetext{a}{Distance from the Sun.}
\tablenotetext{b}{Distance from the Galactic center. The distance of the Sun to the Galactic center is assumed to be 8.5 kpc}
\tablenotetext{c}{Kurayama et al. (2011).}
\label{tab:t1}
\end{deluxetable} 

\clearpage

\begin{deluxetable}{llrrrrrrrr}
\rotate
\tablecolumns{10} 
\tablewidth{0pc} 
\tabletypesize{\small} 
\tablecaption{Observed lines.} 
\tablehead{ 
\colhead{Species} & \colhead{Transition}   & \colhead{Rest Frequency}  & \colhead{$E_u$/$k$} & \colhead{$\mu_0$}  & \colhead{$f_{\rm bw}$\tablenotemark{a}}  & \colhead{$f_{\rm res}$\tablenotemark{b}} & \colhead{$\theta_{\rm FWHM}$}& \colhead{$\eta_{\rm MB}$} & \colhead{$T_{\rm sys}$}\\
 & & [GHz]   & [K] & [D] & [MHz] & [kHz] & [arcsec]&  & [K] \\
 }
\startdata
DNC & $J$= 1--0 & 76.305727 & 3.66 & 3.05 & 40 & 37 & 21 & 0.53 & 250 \\
HN$^{13}$C & $J$= 1--0 & 87.090850 &4.18 & 3.05 & 40 & 37 & 18 & 0.43 & 150 \\
\enddata 
\tablenotetext{a}{The total bandwidth of the backend.}
\tablenotetext{b}{The frequency resolution of the backend.}
\label{tab:t2}
\end{deluxetable} 

\clearpage

\begin{deluxetable}{lrrrr} 
\tablecolumns{5} 
\tablewidth{0pc} 
\tabletypesize{\scriptsize} 
\tablecaption{Line Parameters of HN$^{13}$C $J$=1--0\tablenotemark{a}} 
\tablehead{ 
\colhead{Source}& \colhead{$T_{pk}$} & \colhead{$\Delta V$}   & \colhead{$\int T_a^* dV$} & \colhead{$V_{LSR}$} \\
&\multicolumn{1}{c}{[K]} & \multicolumn{1}{c}{[km s$^{-1}$]} & [K km s$^{-1}$] & [km s$^{-1}$] \\
 }
\startdata 
\multicolumn{5}{c}{MSX \& Spitzer Dark Sources}\\ \hline
G015.05+00.07 MM1  & 0.21(0.01) & 2.13(0.01) & 0.47(0.03) & 24.78(0.04) \\
G015.31-00.16 MM3   & 0.28(0.01) & 0.87(0.05) & 0.22(0.03) & 30.88(0.02) \\
G024.08+00.04 MM3  & 0.15(0.01) & 1.71(0.12) & 0.25(0.03) & 51.78(0.05) \\
G024.08+00.04 MM4  & 0.18(0.01) & 1.21(0.08) & 0.28(0.03) & 52.02(0.03) \\
G034.43+00.24 MM9  & 0.22(0.01) & 1.93(0.10) & 0.39(0.03) & 58.53(0.04) \\ \hline 
\multicolumn{5}{c}{MSX Dark Sources with Spitzer Sources}\\ \hline                      
G019.27+00.07 MM1   & 0.47(0.01) & 2.51(0.05) & 1.37(0.03) & 26.63(0.02) \\
G022.35+00.41 MM1   & 0.27(0.01) & 2.52(0.08) & 0.77(0.03) & 52.70(0.03) \\
G023.60+00.00 MM2  & 0.27(0.01) & 2.71(0.09) & 1.19(0.02) & 53.65(0.04) \\
G034.43+00.24 MM3   & 0.51(0.01) & 1.95(0.03) & 1.19(0.02) & 59.67(0.01) \\
I18151-1208 MM2   & 0.31(0.01) & 2.68(0.08) & 0.98(0.03) & 29.98(0.03) \\
I18223-1243 MM3   & 0.61(0.01) & 2.35(0.03) & 1.50(0.02) & 45.97(0.01) \\
I18337-0743 MM3   & 0.20(0.01) & 1.60(0.07) & 0.34(0.02) & 56.34(0.03) \\ \hline 
\multicolumn{5}{c}{High-Mass Protostellar Objects}\\ \hline                      
I18089-1732 MM1   & 0.44(0.01) & 3.33(0.06) & 1.51(0.02) & 32.80(0.02) \\
I18151-1208 MM1   & 0.43(0.01) & 1.99(0.04) & 1.03(0.03) & 33.37(0.02) \\
I18182-1433 MM1   & 0.35(0.01) & 3.19(0.07) & 1.17(0.03) & 60.16(0.03) \\
I18223-1243 MM1   & 0.39(0.01) & 1.98(0.04) & 0.81(0.02) & 45.35(0.02) \\
I18264-1152 MM1   & 0.50(0.01) & 2.76(0.06) & 1.51(0.04) & 43.61(0.03) \\
I18337-0743 MM1  & 0.40(0.01) & 3.32(0.06) & 1.35(0.02) & 58.44(0.02) \\
\enddata 
\tablenotetext{a}{The numbers in parentheses represent the errors in the Gaussian fitting.}
\label{tab:t3}
\end{deluxetable} 

\clearpage

\begin{deluxetable}{lrrrr} 
\tablecolumns{5} 
\tablewidth{0pc} 
\tabletypesize{\scriptsize} 
\tablecaption{Line Parameters of DNC $J$=1--0\tablenotemark{a}} 
\tablehead{ 
\colhead{Source}& \colhead{$T_{pk}$} & \colhead{$\Delta V$}   & \colhead{$\int T_a^* dV$} & \colhead{$V_{LSR}$} \\
& \multicolumn{1}{c}{[K]} &\multicolumn{1}{c}{[km s$^{-1}$]} & [K km s$^{-1}$] & [km s$^{-1}$] \\
}
\startdata 
\multicolumn{5}{c}{MSX \& Spitzer Dark Sources}\\ \hline
G015.05+00.07 MM1  & 0.12 (0.01) & 1.98 (0.26) & 0.25 (0.04) & 25.00 (0.11) \\
G015.31-00.16 MM3   & 0.13 (0.02) & 1.25 (0.20) & 0.18 (0.05) & 31.15 (0.09) \\
G024.08+00.04 MM3  &  --- &  --- & $<$0.12 &  --- \\
G024.08+00.04 MM4  & 0.11 (0.01) & 1.75 (0.24) & 0.35 (0.04) & 51.71 (0.10) \\
G034.43+00.24 MM9  & 0.19 (0.01) & 2.21 (0.19) & 0.35 (0.05) & 58.55 (0.08) \\ \hline 
\multicolumn{5}{c}{MSX Dark Sources with Spitzer Sources}\\ \hline                      
G019.27+00.07 MM1   & 0.10 (0.01) & 2.22 (0.34) & 0.30 (0.05) & 25.97 (0.14) \\
G022.35+00.41 MM1   & 0.14 (0.01) & 2.76 (0.25) & 0.47 (0.04) & 52.60 (0.10) \\
G023.60+00.00 MM2  &  --- &  --- & $<$0.32  &  --- \\
G034.43+00.24 MM3   & 0.18 (0.01) & 1.18 (0.11) & 0.21 (0.03) & 59.50 (0.05) \\
I18151-1208 MM2   & 0.17 (0.01) & 2.53 (0.20) & 0.59 (0.04) & 29.33 (0.08) \\
I18223-1243 MM3   & 0.23 (0.01) & 2.65 (0.11) & 0.73 (0.03) & 45.75 (0.05) \\
I18337-0743 MM3   & 0.11 (0.01) & 1.78 (0.19) & 0.13 (0.03) & 56.06 (0.08) \\ \hline 
\multicolumn{5}{c}{High-Mass Protostellar Objects}\\ \hline                      
I18089-1732 MM1   & 0.09 (0.01) & 3.39 (0.42) & 0.31 (0.04) & 32.24 (0.18) \\
I18151-1208 MM1   & 0.22 (0.01) & 2.05 (0.14) & 0.45 (0.04) & 33.18 (0.06) \\
I18182-1433 MM1   & 0.16 (0.01) & 2.86 (0.20) & 0.47 (0.04) & 60.26 (0.09) \\
I18223-1243 MM1   & 0.19 (0.01) & 1.43 (0.10) & 0.43 (0.03) & 45.01 (0.04) \\
I18264-1152 MM1   & 0.39 (0.02) & 2.10 (0.10) & 0.91 (0.06) & 43.78 (0.04) \\
I18337-0743 MM1  & 0.14 (0.01) & 3.55 (0.23) & 0.52 (0.03) & 58.83 (0.10) \\
\enddata 
\tablenotetext{a}{The numbers in parentheses represent the errors in the Gaussian fitting.}
\label{tab:t4}
\end{deluxetable} 

\clearpage

\begin{deluxetable}{lrrrrr} 
\tablecolumns{6} 
\tablewidth{0pc} 
\tabletypesize{\scriptsize} 
\tablecaption{Column Density Ratios.} 
\tablehead{ 
\colhead{Source}& \colhead{$N$(DNC)/$N$(HN$^{13}$C)} & \colhead{$N$(DNC)/$N$(HNC)} & \colhead{$T_{\rm rot}$(NH$_3$)\tablenotemark{a}} & \colhead{$T_{\rm k}$\tablenotemark{b}}& \colhead{$N$(HN$^{13}$C)\tablenotemark{c}}\\
& & & \multicolumn{1}{c}{[K]}&\multicolumn{1}{c}{[K]} & 10$^{12}$ cm$^{-2}$ \\}
\startdata 
\multicolumn{5}{c}{MSX \& Spitzer Dark Sources}\\ \hline
G015.05+00.07 MM1 & 0.55$_{-0.11}^{+0.12}$& 0.0111$_{-0.0022}^{+0.0024}$& 15.2$^{+1.7}_{-1.5}$& 19$^{+3}_{-3}$&  2.4\\ 
G015.31-00.16 MM3  & 0.82$_{-0.26}^{+0.30}$& 0.0154$_{-0.0049}^{+0.0055}$& 12.6$^{+2.4}_{-2.0}$& 14$^{+4}_{-3}$& 1.0 \\  
G024.08+00.04 MM3 & $<$0.49& $<$0.0101 & 12.1$^{+1}_{-0.9}$& 14$^{+1}_{-1}$&  1.1 \\   
G024.08+00.04 MM4 & 1.26$_{-0.20}^{+0.24}$& 0.0262$_{-0.0042}^{+0.0051}$& 12.9$^{+1.1}_{-0.9}$& 15$^{+2}_{-1}$& 1.3 \\   
G034.43+00.24 MM9 & 0.92$_{-0.15}^{+0.17}$& 0.0147$_{-0.0024}^{+0.0028}$& 13.9$^{+1.1}_{-0.9}$& 16$^{+2}_{-1}$& 1.9 \\   
\hline 
Average &0.89(0.29)\tablenotemark{d} &0.0169(0.0065)\tablenotemark{d} &13.3(1.2) &15.6(2.0) & 1.5(0.6) \\
\hline
\multicolumn{5}{c}{MSX Dark Sources with Spitzer Sources}\\ \hline                      
G019.27+00.07 MM1  & 0.23$_{-0.04}^{+0.04}$& 0.0041$_{-0.0007}^{+0.0007}$& 15.4$^{+1.1}_{-1.0}$& 19$^{+2}_{-2}$& 6.9\\   
G022.35+00.41 MM1  & 0.62$_{-0.07}^{+0.08}$& 0.0134$_{-0.0015}^{+0.0017}$& 16.0$^{+1.6}_{-1.4}$& 20$^{+3}_{-2}$& 4.0 \\   
G023.60+00.00 MM2 & $<$0.38& $<$0.0081 & 16.0$^{+1.0}_{-0.9}$& 20$^{+2}_{-2}$& 4.4 \\   
G034.43+00.24 MM3  & 0.18$_{-0.03}^{+0.03}$& 0.0029$_{-0.0005}^{+0.0006}$& 15.5$^{+0.9}_{-0.9}$& 19$^{+2}_{-2}$& 6.0 \\   
I18151-1208 MM2  & 0.62$_{-0.06}^{+0.06}$& 0.0116$_{-0.0011}^{+0.0012}$& 17.8$^{+1.5}_{-1.4}$& 24$^{+3}_{-3}$& 5.4\\   
I18223-1243 MM3  & 0.50$_{-0.03}^{+0.04}$& 0.0107$_{-0.0007}^{+0.0008}$& 16.2$^{+1.0}_{-0.9}$& 20$^{+2}_{-2}$& 7.8\\   
I18337-0743 MM3  & 0.39$_{-0.10}^{+0.11}$& 0.0081$_{-0.0022}^{+0.0024}$& 15.0$^{+1.6}_{-1.4}$& 18$^{+3}_{-2}$& 1.7\\   
\hline 
Average &0.42(0.19)\tablenotemark{d} &0.0085(0.0042)\tablenotemark{d} &16.0(0.9) & 20.1(1.7)& 5.2(2.0)\\
\hline
\multicolumn{5}{c}{High-Mass Protostellar Objects}\\ \hline                      
I18089-1732 MM1  & 0.21$_{-0.03}^{+0.03}$& 0.0047$_{-0.0008}^{+0.0007}$& 38 & 194 & 15\\   
I18151-1208 MM1  & 0.46$_{-0.06}^{+0.05}$& 0.0088$_{-0.0011}^{+0.0011}$& 20.8$^{+1.8}_{-1.7}$& 31$^{+5}_{-4}$& 6.4\\   
I18182-1433 MM1  & 0.42$_{-0.04}^{+0.04}$& 0.0105$_{-0.0011}^{+0.0011}$& 19.0$^{+1.5}_{-1.4}$& 26$^{+4}_{-3}$& 6.8\\   
I18223-1243 MM1  & 0.55$_{-0.05}^{+0.05}$& 0.0117$_{-0.0011}^{+0.0012}$& 17.5$^{+1.6}_{-1.5}$& 23$^{+4}_{-3}$&4.5 \\   
I18264-1152 MM1  & 0.62$_{-0.05}^{+0.06}$& 0.0128$_{-0.0011}^{+0.0012}$& 18 & 24 &  8.5\\   
I18337-0743 MM1 & 0.39$_{-0.03}^{+0.03}$& 0.0084$_{-0.0007}^{+0.0007}$& 17.1$^{+1}_{-0.9}$& 22$^{+2}_{-2}$& 7.3 \\   
\hline 
Average & 0.44(0.13)& 0.0095(0.0029)& 21.7(8.1)& 53.4(69.0)&8.1(3.6) \\
\hline
\enddata 
\tablenotetext{a}{The NH$_3$ data are obtained by Sakai et al. (2008), except for I18089-1732 MM1 and I18264-1152 MM1, toward which the NH$_3$ data are reported by Sridharan et al. (2002).}
\tablenotetext{b}{Estimated from $T_{\rm rot}$(NH$_3$).}
\tablenotetext{c}{$T_{\rm rot}$(NH$_3$) is used as the excitation temperature.}
\tablenotetext{d}{Average excluding the upper limits.}
\label{tab:t5}
\end{deluxetable} 

\clearpage

\end{document}